\documentclass[12,fleqn]{elsarticle}
\usepackage{master}
\usepackage{amsmath}
\numberwithin{equation}{section}

\newcommand{\dil}{\phi}	
\newcommand{\bgv}{V}	
\newcommand{\ricci}{\mathcal{R}}
\newcommand{\wsmetric}{\gamma} 
\newcommand{\metric}{g}
\newcommand{\bfield}{b}
\newcommand{\ie}{{\it i.e.}}

\newcommand{\cci}{c_{\text{int}}}
\newcommand{\cce}{c_{\text{ext}}}
\newcommand{\hi}{h_{\text{int}}}

\newcommand{\ns}{\mathrm{NS}}
\newcommand{\ramond}{\mathrm{R}}
\newcommand{\Ein}[1]{\tilde{#1}}

\newcommand{\seff}{S_{\mathrm{eff}}}

\begin{document}

\title{Moduli Stabilisation on the Worldsheet}

\author[kcl]{Jawad Arshad}
\ead{jawad.2.arshad@kcl.ac.uk}
\author[cern]{Neil Lambert\fnref{fn1}}
\ead{neil.lambert@cern.ch}
\author[kcl]{Andreas Recknagel}
\ead{andreas.recknagel@kcl.ac.uk}

\address[kcl]{Department of Mathematics, King's College London, Strand, London WC2R 2LS, United Kingdom}
\address[cern]{CERN Theory Division CH-1211 Geneva 23, Switzerland}
\fntext[fn1]{On leave of absence from King's College London.}

\begin{abstract}
We consider compactifications of type II string theory using exact internal CFT's with central charge $c=9+\epsilon$, $|\epsilon| \ll 1$, leading to an effective potential for the dilaton. For $\epsilon>0$ the potential is positive and the dilaton is ultimately driven to weak coupling. For $\epsilon<0$ the dilaton is driven to strong coupling, but we can stabilise the background by including D-branes. The resulting minimum admits an $AdS_4$ solution where  the cosmological constant is of the  order $\epsilon^3$ and the string coupling constant is of order $\e$. Furthermore these CFT's typically do not possess any massless or tachyonic modes. Thus these vacua provide exact CFT descriptions  of moduli stabilisation in weakly coupled string theory.

\end{abstract}

\maketitle
\newpage

\section{Introduction}

Compactification of the 10-dimensional target space of superstring leads to a vast choice for the data describing the internal compact manifold. Often, especially in supersymmetric compactifications, there are massless moduli fields  that describe the deformations of the internal manifold. A fundamental problem in string phenomenology therefore is to find ways to construct vacua without any moduli since there are strong cosmological constraints on the existence of massless or very light scalar fields. A solution to this problem has been devised over the past few years in the form of flux compactifications (see \cite{Douglas:2006es} for a  review). While they can preserve some supersymmetry, fluxes give vacuum expectation values to the massless fields and hence  stabilise the moduli. According to current wisdom this leads to a large number of vacua  known as the `landscape' since fluxes can take many different discrete values.

These models have been largely studied using effective field theory and supergravity. This requires that the compactification scale is large compared to the string scale but on the other hand is small compared to experimentally accessible scales. From the worldsheet point of view this approach seems somewhat contrived. Therefore it is of interest to consider compactifications of string theory using internal CFT's which do not necessarily  have a large geometric limit. The classic example of CFT compactifications are Gepner Models \cite{Gepner:1987qi}. These are typically thought of as CFT analogues of compact Calabi-Yau manifolds whose size  is of order the string scale, however they are formulated without recourse to geometric data. The construction of Gepner Models  involves taking a tensor product of minimal models whose central charges add up to 9. The  models that we discuss here carry the correct overall central charges $c=15$, but the decomposition differs from the usual $6+9$ split for the internal and external sectors. Instead we take our external theory to be slightly non-flat with $\cce = 6-\e$  and hence $c_{int}=9+\epsilon$, with $\abs{\e} \ll 1$ (other discussions of non-geometric compactifications and moduli stabilisation include \cite{Becker:2006ks}).  We also note that for $\e>0$ similar models were studied long ago within the context of cosmology \cite{deAlwis:1988pr,Antoniadis:1988} , however here we wish to emphasise the issue of  moduli stabilisation.

We will see that there are plenty of explicit internal compact unitary  CFT's  with $|\e | \ll 1$. 
For $\e<0$ there is no limit to how small $\e$ can be. However for $\e>0$  this does not seem to be the case and the best models that we have found have $\e\sim 10^{-6}$.  Keeping $\e$ small means that the theory remains weakly coupled and the external curvatures are small so one recovers a reliable four-dimensional effective field theory.  Furthermore models can be obtained such that the spectrum of the full theory is free of massless fields. The exact spectrum is model-dependent but we discuss general characteristics shared by all models constructed in the prescribed way. Although in this note we will not seek to obtain a realistic spectrum of massless fields, we hope that this can be addressed in future work. 

The rest of this paper is organised as follows. In section 2 we discuss the over-all set-up of our models and derive the four-dimensional effective field theory. We use this to argue that, for $\e<0$ we can introduce D-branes that lead to a stabilising potential for the dilaton. The resulting minimum admits an $AdS_4$
vacuum state at weak coupling and weak curvature. In section 3 we outline the linear dilaton CFT which can be used as an exact background for the external part.  In section 4 we give a brief review of the construction of Gepner models and their D-branes, extending this to the case where the total central charge is $9+\e$. In section 5 we discuss the spectrum of some explicit examples. Section 6 is our conclusion. We also list some examples of the internal CFT's that can be used in the appendix.

\section{Central Charge Deficit and Four-Dimensional Effective Action}

The sigma-model action of the string moving in a background with massless fields is given by,
\begin{align}
	\nonumber
	S = -{1 \over 4\pi\alpha'} \int d^2\sigma \sqrt{\wsmetric} \; \Big\{&
	\wsmetric^{\alpha\beta} \metric_{\mu\nu}(X) \p_{\alpha} X^{\mu} \p_{\beta} X^{\nu} \\
	\label{wsaction}
	&+ \epsilon^{\alpha\beta} \bfield_{\mu\nu} (X) \p_{\alpha} X^{\mu} \p_{\beta} X^{\nu} 
	+\alpha' \ricci \dil(X)\Big\},
\end{align}
where $\alpha,\beta = 0,1$, $\mu,\nu = 0,1,\dots,D-1$, $\ricci$ is the Ricci scalar of the worldsheet metric $\wsmetric_{\alpha\beta}$ . The string coupling constant is $g_s = e^{\dil}$ and hence string perturbation theory is valid where ever $g_s$ is small.  
Although this action is classically conformally invariant this is generically broken in the quantum theory. To one-loop order one finds the conditions for conformal invariance are $\beta^g_{\mu\nu} = \beta^b_{\mu\nu} = 0$ where
\beqs\begin{align}\label{Reqs}
	\beta^g_{\mu\nu}
	&=  \alpha' \Bigl( R_{\mu\nu}-\frac{1}{4}H_{\mu\lambda\rho}H_\nu{}^{\lambda\rho} +2\covdev_\mu \covdev_\nu\phi \Bigr) +\mathcal{O}(\alpha'^2)\\
	\beta^b_{\mu\nu}\label{Heqs}
	&= \alpha' (\covdev^\lambda H_{\mu\nu\lambda} - 2\covdev^\lambda\phi H_{\mu\nu\lambda} ) + \mathcal{O}(\alpha'^2)
\end{align}\eeqs
If these equations are satisfied one finds that the central charge $c$ of the worldsheet conformal field theory is indeed a constant and is given by
\beq
c =  {3\over 2}\beta^\phi
\eeq
where
\beq
	\beta^\phi =D
	+ \alpha'\Bigl( 4(\covdev\phi)^2 - 4\covdev^2\phi -R +\frac{1}{2\cdot3!}H^2 \Bigr)
	+ \mathcal{O}(\alpha'^2).
\eeq
To construct string backgrounds, one needs a worldsheet CFT with $c = 15$. Traditionally this is achieved by taking $D=10$ and solving $\beta^g_{\mu\nu}=\beta^b_{\mu\nu}=0$ and $\beta^\phi=10$ at lowest order. For weakly curved backgrounds, that is curvatures that are large compared to the string scale, this is a good approximation and the resulting equations of motion arise from the supergravity spacetime effective action. Furthermore we wish to take four dimensions to be large, nearly flat, and the remaining ones to be compact. For the $\alpha'$ expansion to be valid however the compact space must be smooth and have a length scale that is large compared with $\sqrt{\alpha'}$.

A way to avoid this is to consider string backgrounds which are a direct product of an internal, exact, CFT with $\cci = 9$ and a non-compact sigma model with a four-dimensional
flat target space and hence $\cce = 6$. A class of exact CFT's with $\cci = 9$ are provided  by the
so-called Gepner models and these are in turn constructed as a
tensor product, with identifications, of minimal models, as discussed below.

Here we will try something slightly different. We will consider the
CFT to be a tensor product of a sigma model with $D=4$ with central charge 6 and an
internal CFT with central charge $\cci =
9+\epsilon$ where $\epsilon$ is a small
dimensionless number, possibly negative. As a result we see that the
conditions for conformal invariance of the non-compact sigma model
are
\beq\label{EOM}
	4-\frac{2\epsilon}{3}
	= 4
	+  \alpha' \Bigl(4(\covdev\phi)^2-4\covdev^2\phi-R+\frac{1}{2\cdot 3!}H^2 \Bigr)
	+ \mathcal{O}(\alpha'^2)
\eeq
where the first equation comes from the condition $c = \cce+\cci=15$. From this equation we see that $\alpha' M^2 \sim \epsilon$, where again $M$ is the scale of the curvature of the background. Provided that $\abs{\epsilon} \ll 1$ then the perturbation expansion is still valid, indeed we can think
of the $\alpha' M^2$ expansion as an expansion in $\epsilon$. We call $\e$ the {\em central charge deficit} following \cite{Antoniadis:1988} and our 
goal will be to search for models with small $\e$. We take the external theory to be in 
$D=4$.

The background equations of motion (\ref{Reqs}),(\ref{Heqs}),(\ref{EOM}) can be derived as the
Euler-Lagrange equations of the spacetime effective action
\beq\label{SFaction}
	\seff = \frac{1}{\alpha'}\int d^4X\sqrt{-g}e^{-2\phi}\left(R +
4(\covdev\phi)^2-\frac{1}{2\cdot 3!}H^2 -
\frac{2\epsilon}{3\alpha'}\right)+\dots,
\eeq
where again the ellipsis denotes higher order terms in $\alpha'$ and
derivatives, \ie\ $\epsilon$. 
To proceed it is useful to go to so-called Einstein frame
\beq
	\Ein{g}_{\mu\nu} = e^{-2\phi}g_{\mu\nu}
\eeq
so that, in terms of $\Ein{g}_{\mu\nu}$ the spacetime action is
\beq
	\Ein{S}_{\mathrm{eff}}
	= \frac{1}{\alpha'}\int d^4X\sqrt{-{\tilde g}}\left(
	{\tilde R} - 2(\covdev\phi)^2
	-\frac{1}{2\cdot 3!}e^{-2\phi} H^2
	- \frac{2\epsilon}{3\alpha'}e^{2\phi }\right)+\ldots
\eeq
where all quantities with tilde are in Einstein frame. Here we find an effective
dilaton potential
\beq
	\Ein{V}(\phi) =\frac{2\epsilon}{3\alpha'} e^{2\phi }
\eeq
We will be interested in cosmological solutions that are consistent with an FRW universe. In this case, it is helpful to dualise he NS-NS Kalb-Ramond field into  an axion $a$ via
\beq
	H_{\mu\nu\lambda} = \varepsilon_{\mu\nu\lambda\rho} e^{2\phi}\partial^\rho a
\eeq
which accounts for another massless modulus.
Thus we have
\beq\label{SF}
	\Ein{S}_{\mathrm{eff}} = \frac{1}{\alpha'}\int d^4X\sqrt{-{\tilde g}}\left({\tilde R} -2(\covdev\phi)^2-\frac{1}{2}e^{2\phi}(\covdev a)^2 -
\frac{2\epsilon}{3\alpha'}e^{2\phi}\right)+\dots,
\eeq

In addition to these universal string fields that arise from the NS-NS ground state of the string, a generic compactification will also lead to other light states, \ie\ scalar field whose mass is small compared to $1/\sqrt{\alpha'}$. Thus the total effective action is
\beq
	\Ein{S} = \Ein{S}_{\mathrm{eff}}  - \frac{1}{2\alpha'}\sum_a\int d^4x \sqrt{-{\tilde g}} (\partial_\mu \chi_a\partial^\mu\chi_a +  e^{2\phi} m_a^2 \chi_a^2)
\eeq
where the $\chi_a$ represent the other light scalar fields and  $ m_a$ is   the string-frame mass, as computed in the CFT. Later in the text, we relate this to the conformal dimension in the test case of an exact background containing linear dilaton in the external, noncompact sector.

\begin{figure}[h!]
	\includegraphics[scale=.45]{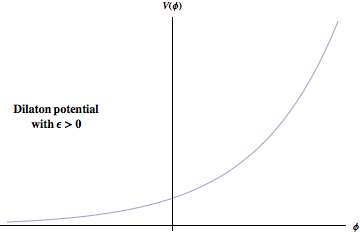}
	\qquad
	\includegraphics[scale=.45]{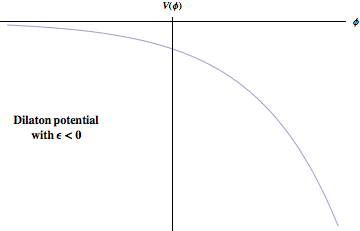}
\end{figure}

For $\epsilon >0$ we find that the dilaton will ultimately run
to weak coupling; $\phi\to-\infty$. However there will be solutions
where $\phi$ initially runs up the potential, stops, and then rolls
back down to $\phi\to-\infty$. At the turning point $V(\phi)>0$ with
$\dot \phi=0$ and hence there will be a short period of inflation as
in the model of \cite{Townsend:2001ea} (only here it is realised
within perturbative string theory).

For large $\phi$ the strings are strongly coupled and $\epsilon$ cannot be used to suppress the curvatures and our $\alpha'$ expansion isn't valid. For $\e>0$ this region can be avoided however if $\epsilon <0$ then we see that eventually $\phi\to\infty$ and we always end up in the uncontrolled regime. 
Suppose that we introduce D-branes into the background which are
extended along the four non-compact spatial directions. Such a brane would introduce a term in the effective
potential of the form
\begin{equation}
S_{\mathrm{brane}}= - T\int d^4x \sqrt{-g}e^{-\phi}
\end{equation}
with $T>0$ and again we are in the string frame. 
Transforming to the
Einstein frame gives
\begin{equation}
\Ein{S}_{\mathrm{brane}}= - T\int d^4x \sqrt{-{\tilde g}}e^{3\phi}
\end{equation}
%

Combining this with the charge deficit contribution we arrive at the effective dilaton potential
\begin{equation}
\Ein{V} (\phi) = \frac{2\epsilon}{3\alpha'}e^{2\phi }+Te^{3\phi}
\end{equation}
This potential has an extremum at
\begin{equation}
\frac{4\epsilon}{3\alpha'}e^{2\phi }+3Te^{3\phi}=0
\end{equation}
which, for positive tension branes ($T>0$), has a solution if
$\epsilon <0$;
\begin{equation}
e^{\phi_0}=-\frac{4\epsilon}{9T\alpha'}
\end{equation}
with
\begin{equation}
\Ein{V}(\phi_0)=\frac{32\epsilon^3}{9^3\alpha'^3T^2}<0
\end{equation}
One can see that this is in fact a stable minimum. For small and negative $\epsilon$ this critical point
is weakly coupled with a small but negative cosmological constant.
\begin{figure}[h!]
\centering
	\includegraphics[scale=.5]{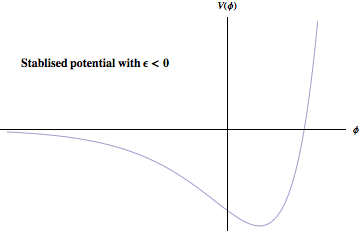}
\end{figure}

In the rest of this paper we will develop  the worldsheet CFT's required in these  constructions. We first consider the case $\e>0$ where the running of the dilaton down the potential can be modeled exactly by a timelike linear dilaton CFT (although there are of course other solution whose exact CFT is unkown). For the $\e<0$ case we need to construct the internal compact CFT and also find suitable D-branes that can be introduced to stabilise the dilaton. We will see that models can readily be found with small $\e$ and no massless fields. In such cases there are no problems with tadpoles from the D-branes and thus they can be consistently incorporated into the background. Furthermore the absence of massless field implies that the only massless moduli left is the axion $a$. However typically non-perturbative such  effects, such as worldsheet instantons, produce a  periodic    potential for $a$ which must therefore have a global minimum. Thus these models have an $AdS_4$ vacuum with no massless fields.

In the following, we will present examples of CFTs that exhibit non-zero
central charge deficits in the external and internal sector. One should keep 
in mind, however, that the non-trivial dilaton potential in principle affects the 
``trustworthiness'' of these CFT backgrounds. For $\epsilon >0$, the dilaton is driven towards
$\phi_0 = -\infty$, thus string loops effects are switched off and the 
tree-level CFT may be regarded as a complete description of the background. 
For $\epsilon>0$, on the other hand, adding a brane makes the dilaton 
settle at a small but finite value of $g_S$; thus the tree-level is not the 
full story, and loop corrections (as well as backreactions of the brane) 
will play a role. At the minimum of the dilaton potential, the theory 
should in principle be described by some new CFT taking these corrections 
into account, and due to the negative cosmological constant $V(\phi_0)$, 
it is natural to suspect this CFT to be realised  by a sigma model with an
$AdS_4$ target space. However a main benefit of our construction is that the derivative expansion turns out to be an expansion in $\e$ and hence the leading order contributions to the effective theory can be very accurate. Thus in these cases we don't expect the exact solution to differ from that obtained from the effective action in any substantial way.

%
%
%
%
\section{The external sector}
%
%
%
%
The external CFTs we will use here to build string backgrounds 
are linear dilaton models. They can be described as sigma models 
with the bosonic part of the action given in 
\eqref{wsaction}, where one sets $\dil(X) = \bgv_{\mu} X^{\mu}$ 
for some fixed vector $V_\mu$ in four-dimensional Minkowski space. 
The main piece of information we will need in the following section 
is the mass shell formula arising from the physical state condition 
in string theory, see below. 

In the supersymetric version, the bosons $X^\mu$ are accompanied 
by free fermions $\psi^\mu$, and the generating fields of the 
(left-moving) $N=1$ superconformal algebra can be written as 
\begin{eqnarray}
    T_{\rm ext}(z) &=& -{1\over \alpha'} \nop{\partial X_{\mu} \partial X^{\mu}} 
 + V_{\mu} \partial^2 X^{\mu}  - \half\, \psi_\mu \partial\psi^\mu,
	\\
\nonumber  G_{\rm ext}(z) & =&  i(2/\alpha')^\half\,\psi_\mu\partial X^\mu -  
          i(2\alpha')^\half\,V_\mu\partial \psi^\mu 
\end{eqnarray}
The central charge of the Virasoro algebra is, in $D=4$ dimensions,  
\beql{ld-cencharge}
	c_{\rm ext} = 6  + 6\alpha' V^2\ ,
\eeq
so unless $V_\mu$ is lightlike, the linear dilaton can serve as the external CFT in 
string compactifications with non-trivial central charge deficit. $V_\mu$ is 
constrained by the choice we will make for $\epsilon$ in the internal CFT, 
\beql{epsilon}
	\epsilon = -6\alpha' V_{\mu} V^{\mu}\ .
\eeq

In many respects, the linear dilaton is close to a free boson CFT, the main 
difference being the $V_\mu$-term in $T_{\rm ext}(z)$. As a consequence, the field 
$\partial X^\mu$ is not a conserved current, instead there is a so-called 
background charge, and correlation functions -- which will not concern 
us in this paper -- would have to be computed using 
screening operators, see \cite{DotsenkoFateev} and also \cite{Antoniadis:1988}.

The fields we are interest in for string theory purposes are the vertex operators
\beq
	\mathcal{V}_p = \nop{\;e^{ip_{\mu}X^{\mu}}} 
\eeq
which are primary with conformal dimension
\beql{vertexopconfdim}
	h_p = {\alpha'\over 4} (\,p_{\mu}p^{\mu} + 2ip_{\mu} V^{\mu})\ .
\eeq

\smallskip

In order to analyse spectra of string compactifications, one starts from the physical 
state condition 
\beq 
(\,L_0^{\mathrm{tot}}-a\,)\;\bra{\mathrm{phys}} = 0
\eeq
to extract a formula for the mass of a string state in terms of conformal 
dimensions. $a$ is a normal-ordering constant. We tacitly assume that 
$\bra{\,\mathrm{phys}}$ is a primary state for the super-Virasoro algebra to 
begin with. 

$L_0^{\mathrm{tot}}$ is a sum of the internal $L^{\rm int}_0$ and of the zero 
mode  of the linear dilaton energy-momentum tensor which acts as 
\beql{lindil-L0}
L_0^{\rm ext} =  {\alpha' \over 4} p_{\mu} p^{\mu} 
 + i {\alpha' \over 2} V_{\mu} p^{\mu} + N \ .
\eeq
Here, the offset $N$ from the conformal dimension \eqref{vertexopconfdim} accounts 
for modes $L^{\rm ext}_n$ or $\psi^\mu_r$ applied to ground states $\bra{p}$; we 
have $N\in \half\,\Z_+$ in the Neveu-Schwarz sector and $N\in\Z_+$ in the Ramond sector. 

Inserting this into the physical state condition, we get 
\beql{phys-state-cond}
	{\al'\over4} p_{\mu} p^{\mu} + i {\al'\over2} V_{\mu} p^{\mu}  = \Delta_{N,a}
	\qquad{\rm with}\qquad \Delta_{N,a} = a - N - \hi
\eeq
and see that we need to admit complex $p^{\mu} = \rho^{\mu} + i\sigma^{\mu}$
in order to have non-trivial solutions -- assuming that the conformal dimensions 
$h_{\rm int}$ from the internal CFT are real. The imaginary part of the equation, 
\beqn
	{\al'\over2}\, \rho_{\mu}\;(\, \sigma^{\mu} +  V^{\mu}\,) = 0,
\eeq
is solved by fixing $\sigma_{\mu} = - V_{\mu}$. Inserting this into the real part 
and using \eqref{epsilon}, we can rewrite the physical state condition in terms of 
the charge deficit $\epsilon$ as 
\beql{phys-state-cond-real}
	{\al'\over4} \rho_{\mu} \rho^{\mu}  = \Delta_{N,a} + {\epsilon \over 24}\ .
\eeq

\smallskip

Some interpretational issues remain. First of all, we have to relate the spacetime 
mass of a string state to the quantities in this physical state condition. 
We will set 
\beql{massformula}
	m^2 \equiv -  {\alpha' \over 4} \big( p_\mu p^\mu + 2i V_\mu p^\mu \big) = - \Delta_{N,a}
\eeq
This choice is motivated by considering the coupling of the linear dilaton to a 
massive scalar field \cite{Chamseddine:1991qu} with the action given by
\beq
	-\frac{1}{2\alpha'}\int d^4 x \sqrt{-g} \; e^{-2\phi} \{ (\p \chi)^2 + m^2 \chi^2 \},
\eeq
whose equation of motion gives the mass-shell condition
\beq
	p_\mu p^\mu + 2i V_\mu p^\mu + m^2 = 0.
\eeq
Thus we can identify $m^2$ in the effective action with the mass-squared computed by the conformal dimension of the internal CFT via (\ref{massformula}). 
By computing quantum fluctuations around the linear dilaton background in the weak 
gravity limit, we can see that the dilaton and the graviton remain massless, even 
when the background affects the physical mass via the background vector $V_\mu$ 
in the physical state condition. 


In the paper \cite{Antoniadis:1988}, the other natural choice 
$m^2 = - {\al'\over4} \rho_{\mu} \rho^{\mu}$ was used, which leads to
an $\e$-dependent mass shift of the whole spectrum due to the presence of 
the linear dilaton. We will briefly come back to this choice in the next 
section, but let us point out right away that the most important qualitative 
feature of our string compactifications does not depend on which of the two 
definitions of spacetime mass we use: either way, we will encounter many 
models that have no massless moduli. Physically the difference arises because the background geometry is not flat and hence the solution to a massive wave equation can behave with a different mass from that which appears in the Lagrangian. 

The second ``interpretational'' issue concerns the normal ordering constant $a$ 
in the physical state condition. Here, we will make the standard choices
\beq
	a_{\ns} = \half, \qquad
	a_{\ramond} = \half - {d\over 16}
\eeq
for the NS and R sectors respectively, where $d/16$ is the contribution coming from 
$d = D-2$ spin fields in the transverse direction. 
This leads to $a_{\ramond} = {3\over 8}$ for our $D=4$ linear dilaton theory.

The work \cite{Myers:1987fv}, on the other hand, introduces an $\epsilon$-shift into 
the normal ordering constant. There the shift arises from solving the light-cone 
gauge constraint, after having introduced non-standard $V_\mu$-dependent  Lorentz 
group generators, which allow to ``restore'' the Lorentz symmetry which is, at the 
face of it, broken by the linear dilaton theory. The effect of the shift in $a$ 
on the physical state condition is as switching between the above definitions of 
mass.

%
%
\section{The internal sector: epsilon-Gepner models}
%
In \cite{Gepner:1987qi}, Gepner introduced algebraic string compacfications 
using exact CFTs instead of sigma-models on compact Calabi-Yau manifolds 
for the internal sector of the worldsheet theory. Gepner's internal CFTs 
are tensor products of minimal models of the $N=2$ super Virasoro algebra, 
such that their central charges add up to $15 - 3D/2$ where $D$ is the 
dimension of the non-compact spacetime. The full theory is then a (GSO-projected) 
tensor product with an external CFT. 

The string backgrounds discussed here, with non-vanishing central charge deficit 
$\epsilon$, are formed in close parallel to Gepner models, therefore we review 
the main building blocks of Gepner's construction. 

The internal sector is comprised of minimal models of the $\susy{2}$ superconformal 
algebra; these are cosets 
\beq
	{SU(2)_k \times U(1)_4 \over U(1)_{2k+4} },
\eeq
where the subscripts denote the levels of the affine Lie algebras; the central charge
is 
\beql{mm-cencharge}
	c(k) = {3k \over k+2}, \qquad  k = 1,2,\dots\ .
\eeq
A minimal model is a rational CFT, whose finitely many irreducible representations
(of the bosonic subalgebra) can be labeled by $(l, m, s)$, where $l$ refers to the 
$SU(2)_k$, $s$ to the $U(1)_4$ in the numerator, and $m$ to the $U(1)_{2k+4}$ in the 
denominator. These labels satisfy
\begin{align}
\nonumber
	&l = 0,1,\dots,k, \qquad
	m = -k-1,-k,\dots,k+2 \qquad
	s = -1,0,1,2,\\
	&\text{with}\quad
	l+m+s \;\text{ even}.
\end{align}
Representations with $s = 0,2$ belong to the NS sector, while those with $s = \pm 1$ to 
the R-sector. Triples $(l,m,s)$ and $(k-l, m+k+2, s+2)$ give rise to the same representation
and are identified. The conformal dimension $h$ and U(1)-charge $q$ of the highest 
weight state with labels $(l, m, s)$ are given by
\beqsl{hq}
\beq
	h^l_{m,s} = {l(l+2) - m^2 \over 4(k+2)} + {s^2 \over 8},  
	\qquad\qquad
	q^l_{m,s} = {m \over k+2} - {s \over 2} 
\eeq
\eeqs
as 
as long as the labels $(l,m,s)$ lie in the so-called {\em standard range}, i.e.\ they satisfy
\beqs
\beq
	l = 0,1,\dots,k, \quad
	\abs{m-s} \le l, \quad
	s = -1,0,1,2,\quad
	l+m+s \text{ even},
\eeq
or
\beq
	l = 1,2,\dots,k,\qquad
	m = -l,\qquad
	s = -2.
\eeq
\eeqs
Every tuple $(l,m,s)$ may be brought into the standard range using the transformations,
\beq
	(l,m,s) \mapsto (l,m+2k+4,s),\qquad
	(l,m,s) \mapsto (l,m,s+4).
\eeq
Representations can be grouped into pairs $(l, m, s)$ and $(l, m, s + 2)$, each such pair
makes up a full $\susy{2}$ super Virasoro module;  all states in the same bosonic 
sub-representation have the same fermion number (or U(1)-charge) modulo two.

A Gepner model is formed from a tensor product of $r$ minimal models with central charges 
$c(k_i)$ together with $D$ external free bosons and fermions. The latter make up an SO$(d)_1$ 
current algebra, where $d=D-2$ (working in the light cone gauge). In Gepner's original 
construction, the levels $k_i$ are chosen such that 
\beq
	c_{\rm int} \equiv \sum_{i=1}^r c(k_i) = 15 - \frac{3}{2}D \ ,
\eeq 
For compactifications to $D=4$ dimensions, we must have $3 < r \le 9$ because 
$1 \le c(k_i) \le 3$; the most famous example is $(3,3,3,3,3)$ which (like many 
other Gepner models) can be related to a sigma model on a Calabi-Yau manifold, 
namely the quintic hypersurface in $\C P^4$.
In the rest of the text, we will occasionally abbreviate the models by collecting 
the $n$ repeated levels together as $k^n$. For instance, $(k_1^3,k_2) 
\equiv (k_1,k_1,k_1,k_2) = (k_1,k_2,k_1,k_1) = \dots$. The quintic therefore may be written as $(3^5)$.
We will also use $c(k_1,k_2,\dots,k_r)$ to refer to the sum $\sum_{i=1}^r c(k_i)$.

To ensure spacetime supersymmetry and tachyon-freedom, the tensor product of minimal 
models and external CFT needs to be subject to certain projections. Most importantly, 
one demands that the total fermion number of any state be odd (the GSO projection). 
We can write this in the form 
\beql{GSOcondition}
	q_{\rm ext} + q_{\rm int} \in 2\Z+1
\eeq
with the sum of the minimal model charges 
\beq
	q_{\rm int} = \sum_{j=1}^r\  {m_j \over k_j + 2} -  {s_j \over 2} 
\eeq
and the charge of the external fermions 
\beql{extcharge}
	q_{\rm ext} = {d\over 2} {s_0 \over 2}\ ; 
\eeq
here, $s_0 = -1,0,1,2$ labels the irreducible representations ($c,o,s,v$) of SO$(d)_1$, 
such that $q_{\rm ext}$ simply amounts to the external fermion number. 
States in the tensor product that do not satisfy \eqref{GSOcondition} are projected out. 

Apart from this GSO-projection, one also has to ensure that only states in tensor 
products of $r+1$ NS or R sectors are packaged together; this is enforced by 
removing all states which do not satisfy 
\beq
	 {d\over 2} {s_0 \over 2} + {s_j\over 2}.\in \Z\ .
\eeq

Projecting out states from a CFT normally spoils modular invariance of the closed 
string partition function -- unless one treats the projection as part of an orbifold 
procedure (more precisely, in the case of Gepner's construction, a simple current 
orbifolding). Modular invariance is then restored by adding suitable twisted 
sectors to the states left from the tensor product theory one started with. 
We refer to \cite{Gepner:1987qi} for more details and for explicit expressions 
of the full partition function. 

\smallskip

In order to have a large-volume interpretation of a Gepner model in terms 
of a sigma model on a Calabi-Yau manifold \cite{Greene:1996cy,Witten:1993yc} 
without fluxes, it is mandatory that $c_{\rm int}$ is a multiple of 3. For the 
purposes of having a consistent string background, however, this restriction 
is dispensable.  
One can instead start from an external SCFT like the linear dilaton for a suitable 
$\epsilon$ and tensor it with a several $N=2$ minimal models giving the 
correct value of $c_{\rm int}$ without violating any of the constraints imposed 
by string theory, provided that that total central charge of internal and external 
sum up to 15,  
\beql{charge-split}
	c = \cce + \cci = 15 \quad{\rm with}\quad
	\cce = 6-\epsilon \quad{\rm and }\quad
	\cci = 9 + \epsilon\ .
\eeq
In particular, the GSO projection can be performed as usual in 
a Gepner models, as long as one replaces $q_{\rm ext}$ from \eqref{extcharge} by 
the fermion number counting the $\psi^\mu$ from the linear dilaton theory. 
The condition that only NS sectors or only R sectors are to be tensorised can, 
of course, also carried over from the $\epsilon=0$ case. Modular invariance 
will be preserved in the same way as in Gepner models, since the projections 
can still be understood as orbifoldings.  


In the examples we have studied, spacetime supersymmetry is broken. Apart from 
that, the main changes when generalising Gepner's construction to non-zero 
charge deficit are the loss of a geometric picture 
involving Calabi-Yau manifolds, and very different spectra of massless and 
light fields, which will be the topic of the next subsection. 

\smallskip
First, however, we have to remember that, for $\epsilon < 0$, we need to add 
a brane to our background in order to keep the dilaton from running away to infinity. 
This brane should fill the external four-dimensional spacetime, i.e.\ satisfy 
Neumann boundary conditions on the external bosons. Branes of that type were 
already studied in \cite{Bershadsky:1991ay} and also in \cite{Li:1995ed}. 
Boundary states for Gepner models were constructed in 
\cite{Recknagel:1997sb,Recknagel:2002qq}; the internal part of these takes the form 
\beql{bdystatecoeffs}
	\bra{\alpha}  =  
	\sum_{\boldsymbol{\lambda},\boldsymbol{\mu}}\ 
	B^{\boldsymbol{\lambda}, \boldsymbol{\mu}}_\alpha
	\bra{\boldsymbol{\lambda}, \boldsymbol{\mu}}\!\rangle
\eeq
with 
\beq
	B^{\boldsymbol{\lambda}, \boldsymbol{\mu}}_\alpha
	\sim  
	\prod_{j=1}^r\  
	{\sin \pi {(l_j+1)(L_j+1)\over k_j+2} \over \sin \pi {l_j+1\over k_j+2}}
	\; e^{i\pi m_jM_j/(k_j+2)} \; e^{-i\pi s_jS_j/2}.
\eeq
where the $\boldsymbol{\lambda}, \boldsymbol{\mu}$ stand for the 
collection of $(l_i,m_i,s_i)$ labels, while $\alpha$ is short for 
the integer $(L_i,M_i,S_i)$ labels of the boundary states themselves. 
These formulas were written with $\epsilon=0$ Gepner models in mind, but with 
the above adaption of the projections, they work just as well in our more 
general situation. The brane we require for $\epsilon < 0$ can therefore 
be any (GSO-projected) tensor product of a Neumann brane for the linear 
dilaton with a boundary state for the internal sector. 

There is one further condition we have to keep in mind, namely that the brane 
we use should not lead to tadpoles for massless RR-fields (tadpoles for massive fields are harmless and are cured by a suitable shift in their vacuum expectation value \cite{Dine:1987bq}). Such tadpoles 
can be cancelled forming suitable (model-dependent) superpositions of the 
above boundary states (exploiting the fact that the coefficients \eqref{bdystatecoeffs} 
contain roots of unity); cf.\ the construction of branes carrying torsion
charges only for the quintic given in \cite{Brunner:2001eg}. 
However, as it turns out, many of the $\epsilon\neq0$ models we studied so far do 
not contain any massless RR-fields, so there are no tadpoles to cancel.

\section{Some examples, and spectra of light fields}

In  this section, we discuss some specific examples, with 
special focus on the content of massless and light fields. We will find pronounced 
differences between cases with non-vanishing central charge deficit and 
Gepner's original construction. 

In a  Gepner compactification to $D=4$ dimensions (with $\epsilon=0$, the number $r$ 
of minimal models used  satisfies $3 < r \le 9$ because $1 \le c(k_i) \le 3$.  
The most famous example is $k=(3,3,3,3,3)$ which (like many other Gepner models) 
can be related to a sigma model on a Calabi-Yau manifold, in this case the quintic 
hypersurface in $\C P^4$. Gepner models typically have quite a large number of massless 
moduli in the NS-sector, coming from the chiral and anti-chiral primaries. 
These are $N=2$ primary fields $|\psi\rangle$ obeying $G^+_{-{1\over2}}|\psi\rangle =0$ 
or $G^-_{-{1\over2}}|\psi\rangle =0$, and it can be shown that they arise from 
tensor products of minimal model (anti-)\-chiral fields, labeled $(l_i, \pm l_i,0)$,  
such that their conformal dimensions sum up to ${1\over2}$. The Gepner model 
corresponding to the quintic, e.g., has 101 + 1  massless moduli of the (chiral,chiral) or 
(chiral,anti-chiral) type. 

Basically, the reason that so many $h_{\rm int}={1\over2}$ fields survive the GSO 
projection is that $c_{\rm int} = 9$: this forces the levels $k_i$ of the 
constituent minimal models -- or more precisely the numbers $k_i+2$ --  to 
satisfy conditions on their relative divisibility. The same denominators 
$k_i+2$ from the minimal model central charges also occur in the 
U(1)-charges $q^l_{m,s}$, and due to the tuned divisibility there are many 
possible choices of $(m_i,s_i)$ satisfying the GSO condition. 
In view of this connection, one might already expect that our epsilon-Gepner
models with $c_{\rm int} = 9+\epsilon$ for $\epsilon \neq 0$ will typically have 
substantially fewer massless modes than Gepner models, or indeed none at all. 
This expectation is confirmed when studying concrete examples.

\smallskip

Before presenting a few of these, let us make some general observations on 
tensor products of $N=2$ minimal models with $c_{\rm int} = 9 + \epsilon$. To get 
$0 < \epsilon \ll 1$, one only needs to consider models with $r=4,\ldots,8$ 
tensor factors; for $r=9$, the smallest possible central charge deficit is 
$\epsilon={1\over2}$, for the model with levels $k=(1^8,2)$.

A small non-zero value of $|\epsilon|$ usually requires at least 
one of the levels $k_i$ to be rather large, which implies that the $i^{\rm th}$ 
minimal model contributes a very large number of fields to the tensor product 
theory; checking the spectrum for light fields is therefore a (somewhat non-trivial) 
task for the computer. 

Having one large $k_i$ also means that there is a narrow mass gap 
$\Delta m^2 \sim (k_i+2)^{-1}$ above the lowest-lying states. In  
\cite{deAlwis:1988pr}, this was viewed as a problematic issue, as low mass gaps 
might be interpreted as a signal for large internal ``dimensions'' of the 
compactification. However, the same feature occurs in ordinary Gepner models 
with $c_{\rm int}=9$. There is no reason to perform a limit 
$\epsilon \to 0$, as is hinted at in \cite{deAlwis:1988pr}; rather, $\epsilon$ is 
a fixed (discrete) input parameter specifying the string background. 
Still, the narrow mass gaps present the following challenge: For phenomenological 
applications, one would prefer models with as vast a ``desert'' above the 
lowest-mass states as possible, so one is faced with an optimisation problem 
balancing small $\epsilon$ against large $k_i$. 

How small one can make $\epsilon$ depends crucially on its sign. 
For $\epsilon < 0$, one can approach $c_{\rm int} = 9$ arbitrarily closely from 
below: e.g., one can start from a tensor product with central charge 6 (from a 
K3 Gepner model), then tensor with one further $N=2$ minimal model, with very 
large $k_i$. We will however also present an example manufactured from a 
$c_{\rm int}=9$ Gepner model below. 

For $\epsilon > 0$, the situation is very different. When restricting ourselves 
to tensor products of $N=2$ minimal models as outlined above, there is actually 
a smallest possible positive $\epsilon$ that can be achieved. The reason is that
the minimal model central charges \eqref{mm-cencharge} form a discrete series between 
1 and 3. More details on series of $c_{\rm int}>9$ for our epsilon-Gepner models 
will be given in the Appendix. 

\smallskip

The ``best'' $c_{\rm int} >9$ model, i.e.\ the one with the smallest positive $\e$ we 
can construct in this way, has levels $k=(1,5,41,1805)$ and central charge
\beq
	c_{\rm int}   = {4895164 \over 543907},
	\quad
	\epsilon\approx  1.8 \times 10^{-6}.
\eeq
Using the mass formula \eqref{massformula} worked out in the previous section, one 
can compute the spectrum of tachyonic, 
massless, and ``light'' fields in the NS and the R sector for this compactification. 
One finds 
\bit
\item Tachyons:        0 (NS), 0 (R). 
\item Massless fields:  1 (NS), 0 (R). 
\item Massive fields with $m^2 < \half$:   161 (NS), 4 (R).
\eit
The single massless state comes from the identity $[(0,0,0),\dots,(0,0,0)]$ in the 
internal minimal model and has external oscillator number $N=\half$, i.e.\ it gives 
the graviton. The next lightest field has $[(0,0,0),(0,0,0),(0,0,0),(0,0,0),(2,0,0)]$ 
and $N=\half$ and has $m^2 = 2/1807 \approx  10^{-3}$, i.e.\ its mass sits at 
about 1000 times the scale set by $\e$. The four massive Ramond-Ramond fields listed 
here are all rather heavy with $m^2 \sim 1/2$. 
Just as in the model considered in \cite{deAlwis:1988pr}, the spectrum 
has no spacetime supersymmetry. 

\smallskip 

A comment on finding such spectra is in order. Since our models typically have one 
level $k_i$ which is very large, the tensor product of minimal models contains very 
many fields -- about $10^{12}$ in the above example $k=(1,5,41,1805)$, 
Sifting through those for suitable $h$-values to identify tachyonic and massless 
fields takes too long a time even on a computer. Therefore, one first reduces the 
number of fields by performing the GSO projection (an easier task since it only 
involves the $m_i$ and $s_i$ quantum numbers, not the $l_i$) and then checks the 
conformal dimensions of the surviving states. 

The problem of listing states becomes even more severe for conformal dimensions
$h_{\rm int}>\half$, as additional $N=1$ primary fields can arise from linear 
combinations of descendants of $N=2$ primaries. As long as we concentrate on 
light string modes, we can however ignore those additional states.

\smallskip

We see that our ``best'' $\e>0$ model has no massless moduli at all. This may be
compared to the closely related model $(1,5,41,1804)$, which  has $c_{\rm int}=9$ 
and a geometric  interpretation as a Calabi-Yau sigma model. Here, one finds 
504  massless moduli, see \cite{Lynker:1990gh,Fuchs:1989pt}.

From this Calabi-Yau Gepner model, one can of course also obtain a compactification 
with small negative central charge deficit: The model with $k=(1,5,41,1803)$ has 
\beq
	c_{\rm int} = {4889744 \over 543305},
	\quad
	\epsilon \approx -1.8 \times 10^{-6}.
\eeq
In this case, one finds the following spectrum of low-lying states: 
\bit
\item Tachyons:         0 (NS), 0 (R).
\item Massless fields:  1 (NS), 0 (R).
\item Massive fields with $m^2 < \half$:   161 (NS), 0 (R).
\eit
Again, the only massless field in the NSNS sector corresponds to the graviton, and 
the next NS lightest field is still $[(0,0,0),(0,0,0),(0,0,0),(0,0,0),(2,0,0)]$ 
with $N = \half$; its mass is $m^2 = 2/1805 \approx 10^{-3}$. 

\smallskip
While the numbers of light fields in the NS and R sector depends crucially on 
the levels $k_i$ defining the compactification, we can make a few general remarks, 
exploiting the GSO projection and familiar bounds on conformal dimensions in $N=2$ 
SCFTs, namely $h_{\rm int} \geq {|q_{\rm int} |\over2}$ in the NS sector and 
$h_{\rm int} \geq {c_{\rm int} \over24}$ in the R sector. This allows to show 
that, with the mass formula \eqref{massformula}, models with $\epsilon > 0$ 
are tachyon-free and moreover have no massless fields in the Ramond sector. For models 
with $\epsilon<0$, one can at least show that there are no tachyonic modes in 
the NS sector. In case all the $k_i+2$ are pairwise coprime, one can make some 
further statements, e.g.\ concerning absence of R ground states from the 
GSO-projected theory. 




Some further examples of models with small central charge deficits are listed 
in the Appendix, along with counts of the low-lying states.

\section{Conclusions and Open Problems}

In this paper we have discussed worldsheet CFTs where the split in central charge between the internal and external sector is shifted to $6-\e$ and $9+\e$. This leads to an effective potential for the  dilaton which can be stabilised by the addition of suitable D-branes. The result is an effective description, valid for small $\e$ that admits a stable $AdS_4$ vacuum with no massless moduli. We presented explicit compact CFTs in the form of epsilon-Gepner models and also some exact D-brane boundary states. In a sense one might think of these epsilon-Gepner models as CFT analogues of Calabi-Yau compactifications with fluxes. Indeed this raises the possibility that in some cases it might be possible to realise a geometrical limit where the epsilon-Gepner models can be related to Calabi-Yau compactifications with fluxes.

Our models in this sense provide a worldsheet analogue of the flux compactification mechanisms that have been studied in supergravity. The advantages of our construction are that the internal CFT can be described exactly and yet one still obtains a spacetime effective picture that is weakly curved and weakly coupled. On the other hand we have not addressed some important questions. These include combining the techniques used here with models that have a realistic spectrum. In addition one would ultimately like to extend the analysis to include a sort of KKLT mechanism \cite{Kachru:2003aw} whereby the cosmological constant is lifted to a positive value. Our models also naively break supersymmetry, already because the D-branes do not carry a charge (and hence cannot saturate a lower bound on the mass by the charge that is characteristic of supersymmetric D-branes). However it could be possible with more care to obtain supersymmetric vacua, or at least vacua where supersymmetry is broken at a small scale determined by $\e$.

We also saw that there are in fact a large number of such models. Indeed since there are infinitely many models with $\e<0$ if we relax the usual Gepner-model constraint that $\e=0$, it would appear that our models will be rather generic since the separate levels of the minimal models are no longer need to be related by having common factors. In such cases the absence of massless moduli seems to be rather generic (but not universal). 
We could also use a variety of other CFTs for the internal sector. For example we could consider tensor products of $N=1$ minimal models, 
which will provide a ``denser set'' of $\epsilon$-values to choose from, while 
still retaining rationality. But other options are available, so there is a 
chance to find models with a more realistic spectrum. In addition, it would be nice to obtain an exact description of the $AdS_4$ vacuum at the minimum of the potential
  
\section*{Acknowledgements}
 
We would like to thank I. Antoniadis, M. Breuning,  I. Brunner, R. Helling, W. Lerche and J. Walcher for discussions. This work was supported in part by STFC rolling grant ST/G000395/1.
 
\newpage

\appendix

\section{`Best' $\e>0$ models}

In the following, a tensor product of $r$ minimal models with levels $(k_1,k_2,\dots,k_r)$ will be written in an ascending order of their levels, i.e., $k_{i-1} \le k_i$. Since $c(k+1) > c(k)$, $c(k'_1,k'_2,\dots,k'_r) \le c(k_1,k_2,\dots,k_r)$ for all $k'_i \le k_i$. The algorithm we use finds the first $(k_1,k_2,\dots,k_r)$, $(k_{i-1} \le k_i)$ model with central change $9+\e$, $\e>0$, i.e., 
\beqn
	c(k_1,k_2,\dots,k_r-1) \le 9 \qquad\text{and}\qquad c(k_1,k_2,\dots,k_r)>9.
\eeq
Each subsequent model $(k_1,k_2,\dots,k_r+1)$ will have a central charge greater than $9+\e$.  Given an upperbound $\e'$, the algorithm also determines the model $(k_1,k_2,\dots,k_{r-1},k_r')$ with highest $k_r'$ such that $c(k_1,k_2,\dots,k_r') \le 9 + \e'$.

In the following table, we list models with $0 < \e \le 0.05$. The total number of models in this range is infinite and we list a few series of models from $(k_1,k_2,\dots,k_r)$ to $(k_1,k_2,\dots,k'_r)$ as defined above.  We find the total number of these series to be 160.
\begin{center}
\begin{table}[!h]
\caption{{\bf Best $\boldsymbol{\e > 0}$ models}}
{\small
\begin{tabular}{|c|c|c|c|c|}
	\hline
	$r$ & $(k_1,k_2,\dots,k_r)$ & $c$ & $\e$ & $(k_1,k_2,\dots,k_r')$\\
	\hline
	4 & (1,5,41,1805) &  4895164/543907 & $1/543907 \sim 1.84 \times 10^{-6}$ & $(1,5,41,\infty)$ \\
	&(1,5,42,923) & 1282051/142450 &  $1/142450\sim 7 \times 10^{-6}$  & $(1,5,42,\infty)$ \\
	&(1,5,45,393) & 1169596/129955 & $1/129955\sim 7.7 \times 10^{-6}$  & $(1,5,45,\infty)$ \\
	&(1,5,43,629) & 596296/66255 & $1/66255 \sim 1.5 \times 10^{-5}$  & $(1,5,43,\infty)$ \\
	&(1,5,44,482) & 350659/38962 & $1/38962\sim 2.5\times 10^{-5}$ & $(1,5,44,\infty)$ \\
	& (2,3,19,419) & 265231/29470& $ 1/29470 \sim 3.4 \times 10^{-5}$ & $(1,5,19,\infty)$ \\
	& (3,3,9,109) & 18316/2035 & $ 1/2035 \sim 4.9 \times 10^{-4}$ & (3,3,9,1318)\\
	& (4,4,5,41) & 2710/301 & $1/301 \sim 3.32 \times 10^{-3}$ & (4,4,5,62) \\
	& (1,5,86,86) & 1387/154 & $1/154 \sim 6.5 \times 10^{-3}$ & (1,5,86,241) \\
	&(1,5,87,87) & 5612/623 & $5/623\sim 8.03\times 10^{-3}$ & (1,5,87,233) \\
	& (5,5,7,8) & 947/105 & $ 2/105 \sim 1.9 \times 10^{-2}$ & (5,5,7,8) \\
	\hline
	5& (1,1,2,11,155) & 36739/4082 & $1/4082 \sim 2.45 \times 10^{-4}$ & $(1,1,2,11,\infty)$ \\
	&(1,1,2,15,39) & 12547/1394 & $1/1394\sim 7.2 \times 10^{-4}$ & (1,1,2,15,59) \\
	& (1,1,2,12,83) & 10711/1190 & $1/1190\sim 8.4 \times 10^{-4}$ & (1,1,2,12,278) \\
	& (1,1,2,13,59) & 5491/610 & $1/610 \sim 1.64 \times 10^{-3}$ & (1,1,2,13,118) \\
	& (2,2,2,4,11) & 235/26 & $1/26 \sim 3.85 \times 10^{-2}$ & (2,2,2,4,11) \\
	\hline
	6& (1,1,1,1,5,41) & 2710/301 & $1/301 \sim 3.32 \times 10^{-3}$ & (1,1,1,1,5,62) \\
	& (1,1,1,1,6,23) & 901/100 &  $1/100 = 10^{-2}$ & (1,1,1,1,6,28) \\
	& (1,1,1,1,7,17) & 514/57 & $1/57 \sim 1.75 \times 10^{-2}$ & (1,1,1,1,7,19) \\
	\hline
	7& (1,1,1,1,1,2,11) & 235/26 & $1/26 \sim 3.85 \times 10^{-2}$ & (1,1,1,1,1,2,11) \\
	& (1,1,1,1,1,3,6) & 181/20 & $1/20 = 5\times 10^{-2}$ & (1,1,1,1,1,3,6) \\
	\hline
\end{tabular}
}
\label{best-pos-eps-models}
\end{table}
\end{center}

Note that the model (1,5,87,86) has a smaller $\e$ than that of (1,5,87,87) but since we order models in ascending order of their levels, $(1,5,87,86)$ is included in the $(1,5,86,86)$ to $(1,5,86,241)$ series. Also note that there is no $r=8$ model with $\e \le 9.05$ as the first model after $c(1^7,4) = 9$ gives $c(1^7,5) \sim 9.14$.

The first model in the table $(1,5,41,1805)$ has the smallest possible positive $\e \sim 1.84 \times 10^{-6}$. Note that $c(1,5,41,1804) = 9$. The fact that $(1,5,41,1805)$ has the smallest $\e$ can be shown by starting with the model $(1^4) = (k_1 = 1,k_2 = 1,k_3=1,k_4=1)$ and incrementing $k_i$ one at a time resulting in the smallest step increase in the central charge. If $k_i$, $i=1,2,3$ are kept fixed at 1 and $k_4$ is increased, we will never obtain a total central charge of 9. Hence we find a bound $k_3 > 1$. By induction on this argument, we see that at least 3 $k_i$'s must be greater than 1. We repeat this procedure until the combination $(1,5,41,1804)$ is obtained with central charge 9 and the very next model $(1,5,41,1805)$ provides us with the smallest possible change in the central charge over 9. We can also see that there are infinite number of $r=4$, $c\le9.05$ models with three levels $k_1 = 1,k_2=5,k_3=41$ by observing that $c(1,5,41) < 6 .05$ and $c(k)$ is bounded from above by 3.

\section{Examples of models}
Here we summarise data calculated for a few example models. The mass scale for light fields is set to $M^2 = 1000\e$ in all cases below.
\begin{center}
\begin{table}[!h]
\caption{{\bf Spectrum of some $\boldsymbol{\e > 0}$ models}}
{\small
\begin{tabular}{|c|c|c|c|c|c|}
	\hline&&&&&\\
	{\bf Levels} &
	{\bf Central} &
	{\bf Charge} &
	{\bf Tach-} &
	{\bf Massless} &
	{\bf Light fields} \\
	$k_i$&{\bf charge} $c$ & {\bf deficit} $\e$& {\bf yons} & {\bf fields}& ($m^2\le M^2$)\\
	\hline&&&&&\\
	(1,5,41,1805) & 4895164/543907 & $ 1.84 \times 10^{-6}$
	&0 	& 1 (NS), 0 (R)	& 1 (NS), 0 (R)\\
	(1,5,41,1806) & 2448937/272104 & $ 3.68 \times 10^{-6}$
	&0 	& 1 (NS), 0 (R)	& 2 (NS), 0 (R)\\
	(1,5,42,923) & 1282051/142450 &  $7 \times 10^{-6}$
	&0 	& 1 (NS), 0 (R)	& 2 (NS), 0 (R)\\
	(1,5,45,393) & 1169596/129955 & $7.7 \times 10^{-6}$
	&0 	& 1 (NS), 0 (R)	& 1 (NS), 0 (R)\\
	(1,5,43,629) & 596296/66255 & $1.5 \times 10^{-5}$
	&0 	& 2 (NS), 0 (R)	&4 (NS), 0 (R)\\
	 (1,5,44,482) & 350659/38962 & $2.5\times 10^{-5}$
	 &0 	& 2 (NS), 0 (R)	&3 (NS), 0 (R)\\
	 &&&&&\\\hline
\end{tabular}
}
\label{pos-eps-models}
\end{table}
\begin{table}[!h]
\caption{{\bf Spectrum of some $\boldsymbol{\e < 0}$ models}}
{\small
\begin{tabular}{|c|c|c|c|c|c|}
	\hline&&&&&\\
	{\bf Levels} &
	{\bf Central} &
	{\bf Charge} &
	{\bf Tach-} &
	{\bf Massless} &
	{\bf Light fields} \\
	$k_i$&{\bf charge} $c$ & {\bf deficit} $\e$& {\bf yons} & {\bf fields}& ($m^2\le M^2$)\\
	\hline&&&&&\\
	(1,5,41,1803) & 4889744/543305 & $-1.84\times 10^{-6}$
	&0 	& 1 (NS), 0 (R)	& 1 (NS), 0 (R)\\
	(1,5,41,1802) & 2443517/271502  & $-3.68\times 10^{-6}$
	&0 	& 1 (NS), 0 (R)	& 2 (NS), 0 (R)\\
	(1,5,42,921) & 1279277/142142 & $-7.04\times 10^{-6}$
	&0 	& 1 (NS), 0 (R)	& 2 (NS), 0 (R)\\
	(1,5,43,627) & 594404/66045 & $-1.51\times 10^{-5}$
	&0 	& 2 (NS), 0 (R)	& 4 (NS), 0 (R)\\
	(1,5,44,480) & 349208/38801 & $-2.58\times 10^{-5}$
	&0 	& 2 (NS), 0 (R)	& 3 (NS), 0 (R)\\
	(1,5,45,392) & 583315/64813 & $-3.1\times 10^{-5}$
	&0 	& 1 (NS), 0 (R)	& 3 (NS), 0 (R)\\
	 &&&&&\\\hline
\end{tabular}
}
\label{neg-eps-models}
\end{table}
\end{center}


\end{document}